\documentclass[12pt,a4paper]{article}

\pdfoutput=1

\usepackage{latexsym}
\usepackage{graphicx}
\usepackage{mathptmx}
\usepackage{bm}
\usepackage{xcolor}
\usepackage{upgreek}
\usepackage{amsmath}
\usepackage{amsfonts}
\usepackage{amssymb}
\usepackage{amsbsy}
\usepackage{amsthm}
\usepackage{dsfont}
\usepackage[T1]{fontenc}
\usepackage[sort&compress,round,comma,numbers]{natbib}

\usepackage[pdftex,colorlinks=true,urlcolor=blue,citecolor=black,anchorcolor=black,linkcolor=black]{hyperref}

\newtheoremstyle{wsc}
{3pt}
{3pt}
{}
{}
{\bf}
{}
{.5em}
{}

\theoremstyle{wsc}
\newtheorem{theorem}{Theorem}

\newcommand{\bb}[1]{\mathbb{#1}}
\renewcommand{\v}[1]{\bm{#1}}
\newcommand{\m}[1]{\textsl{#1}}
\newcommand{\f}[1]{\textit{#1}}

\renewcommand{\c}[1]{\mathcal{#1}}

\begin{document}

%
%

\title{ESTIMATION WHEN BOTH  COVARIANCE
 AND PRECISION MATRICES ARE SPARSE}

\author{Shev MacNamara\\
Erik Schl\"ogl\\[12pt]
School of Mathematical \& Physical Sciences\\
University of Technology Sydney\\
Broadway, Ultimo NSW 2007\\
Sydney, AUSTRALIA\\
~\\
~\\
~\\
\and Zdravko Botev\\[12pt]
School of Mathematics \& Statistics\\
The University of New South Wales\\
High St. Kensington, NSW 2052\\
Sydney, AUSTRALIA
}

\maketitle

\section*{ABSTRACT}
We offer a method to estimate a covariance matrix in the special case that \textit{both} the covariance matrix and the precision matrix are sparse --- a constraint we call double sparsity. The estimation method is maximum likelihood, subject to the double sparsity constraint. In our method, only a particular class of sparsity pattern is allowed: both the matrix and its inverse must be subordinate to the same chordal graph. Compared to a naive enforcement of double sparsity, our chordal graph approach exploits a special algebraic local inverse formula. This local inverse property makes computations that would usually involve an inverse (of either precision matrix or covariance matrix) much faster. In the context of estimation of covariance matrices, our proposal appears to be the first to find such special pairs of covariance and precision matrices.

\section{INTRODUCTION}
\label{sec:intro}
We begin with a quote from \citet*{Rothman2010} 
 ``\textit{As a rule of thumb in practice, if it is not clear from the problem whether it is preferable to regularize the covariance or the inverse, we would recommend fitting both and choosing the sparser estimate}.'' The authors are describing methods that can estimate a covariance matrix or a precision matrix when one -- but not both -- of these matrices is sparse.
For example, \citet*{bickel2008regularized} exploit sparsity patterns when estimating a covariance matrix or its inverse, and \citet*{ravikumar2011high} explore related problems for high dimensional estimation.

In this article we demonstrate that it is in fact possible to be greedy, and ask for simultaneous sparsity  in both the covariance and precision matrices. Our estimator is a maximum likelihood estimator with the constraint that both the covariance  and its inverse be sparse.
To the best of our knowledge, this is the first proposal to seek efficient estimation for the doubly sparse case (when both covariance matrix  and precision matrix are sparse).

Some of the advantages to imposing sparsity in both the covariance and its inverse are simplicity of interpretation and faster computation via local formulas that we describe below (these formulas allow us to work with both the matrix and its inverse in an efficient way). 

Estimating a covariance matrix is an important problem in the subject of statistics. 
Methods of imposing sparsity in the covariance matrix --- or  more commonly in the inverse of the covariance matrix (the precision matrix) --- have attracted a great deal of attention.
A very popular example of such an approach is the \textit{graphical LASSO} of \citet*{friedman2008sparse}. 
One reason these methods are important is that the corresponding model is more  interpretable when the precision matrix is sparse, which is important in the subjects of Gaussian Markov
random fields \cite{l2019gaussian} and of graphical models and covariance selection \cite{GraphicalModelsLauritzenBook}.


When there is a known relationship between entries in a matrix (covariance matrix, precision matrix, or Cholesky factor, for example) and coefficients in regression, then it is possible to apply many available methods from regression that impose zeros in the regression coefficients, as a way to impose zeros in the matrix. Financial applications of such methods    include the shrinkage estimation algorithm of \citet*{ledoit2004honey} and
the missing data completion algorithm of \citet*{georgescu2018explicit}.

Missing data completion is closely related to the \textit{Dempster completion}, which is closely related to the local inverse formula that we describe later.
The  \citet*{Dempster72} completion of a covariance matrix in which there are missing entries has an especially satisfying form when the inverse is chordal.
Then \citet*{georgescu2018explicit} note the completion brings together a number of attractive properties  
\cite{StrangMacNamaraLocalInverseFormula,grone1984positive,dym1981extensions,JohnsonLocalInverse1998}: 
\begin{itemize} 
\item sub-blocks of these matrices away from the main diagonal have low rank, as in The Nullity Theorem \cite{strang2010fast,strang2004interplay} ; these matrices are examples of semi-separable matrices \cite{SemiSeparableBook};
\item the inverse can be found directly, without completing the covariance matrix, via a `local inverse formula' (as  shown in the symmetric positive definite case by \cite{GraphicalModelsLauritzenBook,SpeedKiiveri1986}) that uses only information in the blocks on the main diagonal of the incomplete covariance matrix; 
\item the completed matrix maximizes the $\log$-determinant amongst the cone of symmetric positive definite matrices consistent with the original incomplete covariance matrix, and it is a maximum entropy estimate.
\end{itemize}

When an existing method succeeds in imposing sparsity in the precision matrix,  then typically the corresponding covariance matrix is not sparse.
\textit{Vice-versa}, if the covariance matrix is sparse then the precision matrix is typically not sparse.
In summary, all methods that are currently available in the literature are  \textit{not} able to impose sparsity simultaneously in \textit{both} the covariance matrix and the precision matrix.
Indeed, informally, if a sparse matrix is chosen ``at random'' then \textit{all} entries of the corresponding inverse matrix are non-zero.
A sparse matrix with a sparse inverse is an extremely exceptional case.
That exceptional case, applied to the problem of estimating a covariance matrix, is the subject of the subsequent sections.

In all subsequent sections, we refer to a chordal graph $\c G$ and the junction tree $\c J$ for that graph.
A definition of a chordal graph is that all cycles of four or more vertices have a chord. 
A chord is an edge that is not part of the cycle, but that connects two vertices of the cycle.
There are other equivalent characterizations of chordal graphs, such as the graphs that have perfect elimination orderings.
(See, e.g. \cite{vandenberghe2015chordal,StrangMacNamaraLocalInverseFormula,JohnsonLocalInverse1998}.)
Chordal graphs are also known as decomposable graphs, in the graphical models literature \cite{GraphicalModelsLauritzenBook}.

\section{DOUBLY SPARSE COVARIANCE AND CONSTRAINED LIKELIHOOD}
Block diagonal covariance matrices are the simplest examples for which both the matrix and its inverse are sparse.
A diagonal covariance matrix $\m V = \textrm{diag}(\sigma_1^2, \ldots, \sigma_n^2)$   has inverse $\m V^{-1} = \textrm{diag}(1/\sigma_1^2, \ldots, 1/\sigma_n^2)$.
Similarly, a block-diagonal matrix and its inverse  have the same sparsity pattern. 

Other than such (block) diagonal trivial examples, it is not clear if it is possible 
to construct nontrivial examples in which both the covariance and its inverse have the same sparsity pattern --- a phenomenon we dub \emph{doubly sparse covariance}. 


Do such doubly sparse covariance matrices exist?
The answer is `yes', and examples can be constructed, where both the covariance matrix and its inverse (or precision matrix) are assumed chordal:
\begin{equation}
\label{eq:first:example:V}
\m V \equiv
\left(
\begin{tabular}{cccccc}
    13  &   8  &   4  &   2  &   0 &    0 \\
     8   & 13  &   2  &   1  &  0   &  0 \\
     4   &  2   & 10  &   6  &   1   &  1 \\
     2   &  1   &  6   & 13  &  10  &  10 \\
     0   &  0   &  1   & 10  &  13  &   8 \\
     0   & 0    & 1    & 10  &   8   & 13 
\end{tabular}
\right) 
\end{equation}
with inverse (or precision matrix)
$
\Theta \equiv \m V^{-1} 
$
given by
\begin{equation}
\label{eq:first:example:Vinv}
\Theta  =
\left(
\begin{tabular}{rrrrrr}
        2960  &     $-1690$  &      $-900$    &      90       &    0        &   0 \\
       $-1690$   &     2675  &       150    &     $-15$       &    0        &   0 \\
        $-900$     &    150    &    8715     & $-12180$     &   5385   &     5385 \\
          90     &    $-15$     & $-12180$     &  23835      &$-10770$   &   $-10770$ \\
           0      &     0       & 5385      &$-10770$        &7539      &  3231 \\
           0      &     0       & 5385      &$-10770$       & 3231      &  7539 
\end{tabular}
\right) / 21540 .
\end{equation}
It can be quickly checked this pair are also positive definite.
Both $\m  V$ and $\Theta$ are \textit{not} block diagonal, demonstrating that the class of matrices we are considering in this article is richer than simply block diagonal matrices.

While chordal covariance matrices have been studied before, there seem to be no examples in the literature where both the covariance  and its inverse are chordal.
We will revisit this same matrix example later in \eqref{eq:local:formula:revisit:example} to show that it has some local inverse properties.

\subsection{Constrained Maximum Likelihood}

We want both the matrix $\m M$ and the inverse $\m M^{-1}$ to be \textit{subordinate} to the same chordal graph.
 i.e. if there is no edge between nodes $i$ and $j$ then the $(i,j)$ entry of $\m M$ is zero, and  the $(i,j)$ entry of $\m M^{-1}$ is zero.
We assume that we are given iid sample data $\v X_1,\ldots,\v X_n\in \bb R^{p}$, drawn from $\mathcal N(\v\mu,\m M)$ for some unknown $\v\mu$ and $\m M$.
Let the sample covariance matrix be
\[
\m S = \frac{1}{n} \sum_{i=1}^n (\v X_i - \hat{\v \mu}) (\v X_i - \hat{\v \mu})^\top,\quad 
\hat{\v \mu} = \frac{1}{n} \sum_{i=1}^n \v X_i.
\]
Under the normality assumption and after eliminating the nuisance parameter  $\v\mu$ (by replacing it with its MLE $\hat{\v \mu}$), the maximization of the  log-likelihood  is equivalent to the minimization (with respect to $\m M$):
\[
\mathrm{trace}(\m S\m M^{-1})+\ln|\m M|.
\]
In our constrained MLE  framework, we want to find a symmetric positive definite matrix $M$ that \textit{minimizes the objective function}
\begin{equation}
\textrm{trace} \left(\m M^{-1} \m S \right) + \ln|\m M|
\label{eq:log:likihood:normal}
\end{equation}
\textit{subject to the constraint} that 
\begin{equation}
\m M \textrm{ and } \m M^{-1} \;\; \textrm{ are subordinate to} \;\; \c G.
\label{eq:naive:constraint}
\end{equation}


The main novelty in our constrained optimization is that we will make use of the so-called \emph{local inverse formula} to impose the doubly sparse constraint in \eqref{eq:naive:constraint}. We describe that formula in detail in the next section.
This is a simple formula that relates the  inverse to inverses of sub-blocks of the matrix, and which applies when the graph is subordinate to a chordal graph.

The same idea allows us to compute $\det(\m M)$ in the objective function by a local formula for the determinant \citet{JohnsonLocalInverse1998}.
A main advantage is that we can easily define a function $\f{L}(\m M)$ based on the local inverse formula that allows us to parameterize the set of matrices in the constraint using only one subset of the variables. 
For example, we can use only the subset of nonzero entries of $\m M$, and then $\f{L}(\m M)$ can be used in place of $\m M^{-1}$.
(Or vice-versa: we could parameterize with only the subset of entries of $\m M^{-1}$, and then use $\f{L}(\m M^{-1})$ in place of $\m M$.)

The optimization also exploits the following well-known observations:
\begin{itemize}
\item We need only optimize over a subset of the entries of a Cholesky factor, $\m R$, that correspond to the chordal graph. 
Then form $\m M$, if required, as $\m M = \m R^\top \m R$, for example.
That is, \emph{we parameterize only with the subset of entries in a Cholesky factor}, $\m R$.
\item Parameterizing via a Cholesky factor automatically ensures that we optimise over positive definite matrices. Also, parameterizing via a Cholesky factor  exploits our knowledge of numerical analysis and our knowledge of chordal graphs,  that chordal graphs also correspond to `perfect eliminators.'
That is, the sparsity pattern is preserved \cite{BlairPeytonReport1993}.
 \item Sometimes it may be better to parameterize by the entries of the precision matrix, or by the entries of the Cholesky factor of the precision matrix, as in \citet*{PourahmadiCovarianceMonograph}, but we have made no attempt to compare the relative merits of those two approaches.

\end{itemize}

Before proceeding with the details of the local inverse formula, we make two remarks. 

\paragraph{Imposing the double sparse constraint in a naive way.}
A naive approach to \eqref{eq:log:likihood:normal}+\eqref{eq:naive:constraint} is to apply off-the-shelf optimization software, that only optimizes over the subset of the entries in $\m M$ that are allowed to be possibly nonzero (so that the first part of the constraint, $\m M$ subordinate to $\c G$, is automatically fulfilled), and then impose some form of penalty based on  terms in $\m M^{-1}$ that are nonzero and that are not allowed to be nonzero according  to the graph $\c G$.
There are other naive ways that one could imagine to impose the doubly sparse constraint in \eqref{eq:naive:constraint}.
An issue with such naive approaches is that they do not  make use of the underlying algebraic structure of such special pairs of matrices.

\paragraph{Separable nonlinear least squares problems.}
Finding a pair of matrices $\m M$ and $\m M^{-1}$ that are both subordinate to the same graph $\c G$, as in \eqref{eq:naive:constraint}, is related to so-called  separable nonlinear least squares problems \cite{golub2003separable}.
Roughly speaking, that class of problems corresponds to a nonlinear optimization problem where the variables can be partitioned into two subsets, and where knowing the values of the variables for one of the subsets, leads to a linear problem to be solved in order to find the unknown values of the remaining variables in the other subset.
For our application at hand, the two subsets are the unknown entries of $\m M$ and of $\m M^{-1}$.  
If we knew the true nonzero entries of $\m M$, then it is a simple linear problem to find the nonzero entries of $\m M^{-1}$ (and vice-versa).
In the context of separable nonlinear least squares problems, one approach is the so called \emph{variable projection method}.
Such an algorithm  alternates between updating the two subsets of the variables.
At first glance, the algorithm only involves  linear optimization in each iteration.
Unfortunately, such an alternating approach often has disappointing performance, and there are challenges with parameterization. 

\subsection{Local Inverse Formula}
If the inverse matrix, $\m V^{-1}$, is subordinate to a chordal graph, then this local inverse formula \eqref{eq:local:inverse:formula} tells us how to find the inverse \emph{using only sub-blocks} of the matrix $\m V$. The formula reads
\begin{equation}
\Theta \equiv \m V^{-1} = \sum_{[c] \in \mathcal{C}} \left(\m V_{[c]}\right)^{-1} - \sum_{[j] \in \mathcal{J}} \left(\m V_{[j]}\right)^{-1},
\label{eq:local:inverse:formula}
\end{equation}
where $\m V_{[c]}$ denotes a square sub-block of the matrix $\m V$, which corresponds to a \textit{maximal clique} $c$ in the set of maximal cliques  $\mathcal{C}$ that are the nodes of the \textit{clique tree} associated with the chordal graph of the matrix $\m V$;
and for each element in the set of edges  $\mathcal{J}$ of the clique tree, $\m V_{[j]}$ is a square sub-block of the matrix $\m V$ that corresponds to the \textit{intersection} (or `separator') $j$ of two maximal cliques that are connected in the clique tree.
Proofs of \eqref{eq:local:inverse:formula} can be found in the references, under mild assumptions \citet{JohnsonLocalInverse1998}.
There are also other terminologies for the same thing, see, e.g., Bartlett's lecture notes on Undirected graphical models: Chordal graphs, decomposable graphs, junction trees, and factorizations \url{https://people.eecs.berkeley.edu/~bartlett/courses/2009fall-cs281a/}, or \cite{GraphicalModelsLauritzenBook,SpeedKiiveri1986}.

For our special class of matrices, that satisfy  the doubly sparse constraint in \eqref{eq:naive:constraint},  both are subordinate to the same chordal graph, so we are allowed to swap the roles of $\m V$ and $\m V^{-1}$ and the local inverse formula \eqref{eq:local:inverse:formula} still holds. We can exploit this local property in whichever way is most convenient.
 This is what distinguishes the class of covariance matrices we study here from the rest of the literature.

 \paragraph{Example: A $5\times 5$ Local Inverse Formula.}
While the formula \eqref{eq:local:inverse:formula} scales to large matrices in a computer code, here we only give small examples that can be displayed on a page.
The numerical conditioning depends on the conditioning of the cliques and the separators.
How well the computations scale depends on the clique tree, and especially the size of the maximal cliques. In other words, the speed  depends more on the graph, rather than the  size of the matrix.

Let us revisit the same example in \eqref{eq:first:example:V}.
We will demonstrate the Local Formula \eqref{eq:local:inverse:formula} holds. In plain English, formula \eqref{eq:local:inverse:formula} roughly states that \textit{``the inverse is the sum of the inverses of the blocks, minus the inverse of the overlaps,''} as in this example:
\begin{equation}
\label{eq:local:formula:revisit:example}
\Theta = \m V^{-1} = 
\left(
\begin{array}{ccc}
\left(
\begin{array}{cccc}
    13  &   8  &   4  &   2  \\
     8   & 13  &   2  &   1 \\
     4   &  2   & 10  &   6  \\
     2   &  1   &  6   & 13
\end{array}
\right)^{-1} & \\
& 
\end{array}
\right)
 + 
 \left(
\begin{array}{ccc}
& \\
& 
\left(
\begin{array}{cccccc}
         10  &   6  &   1   &  1 \\
       6   & 13  &  10  &  10 \\
       1   & 10  &  13  &   8 \\
        1    & 10  &   8   & 13 
\end{array}
\right)^{-1} 
\end{array}
\right)
\end{equation}
\[
-
\left(
\begin{array}{ccc}
& \\
& \left(
\begin{array}{cc}
      10  &   6   \\
      6   & 13  
\end{array}
\right)^{-1} \\
& 
\end{array}
\right).
\]

We are also seeing examples of the sub-matrices that correspond to the two maximal cliques (corresponding to the two sets of  indices $\{1,2,3,4 \}$ and $\{3,4,5,6\}$) and to their intersection (corresponding to the set of  indices  $\{3,4\}$) in the clique tree coming from the chordal graph associated with this matrix example.

Recall that what is novel about the general class of matrices that we consider in this article is that they satisfy that local inverse formula, \eqref{eq:local:inverse:formula}, in \emph{ both} directions.
That is, for these special examples, we can swap the roles of $\m V$ and  $\m V^{-1}$, and the Local Formula \eqref{eq:local:inverse:formula} remains true! In this example we obtain:

\begin{equation}
\label{eq:local:formula:revisit:example:REVERSE}
\m V =  21540
\left(
\begin{array}{ccc}
\left(
\begin{array}{cccc}
        2960  &     -1690  &      -900     &      90        \\
        -1690    &     2675  &       150    &      -15        \\
         -900      &    150    &    8715     &  -12180      \\
          90     &     -15      &  -12180      &  23835      
\end{array}
\right)^{-1} & \\
& 
\end{array}
\right)
\end{equation}
\[
 + 
 21540
 \left(
\begin{array}{ccc}
& \\
& 
\left(
\begin{array}{cccccc}
            8715     &  -12180      &   5385   &     5385 \\
        -12180      &  23835      & -10770    &    -10770  \\
            5385      & -10770         &7539      &  3231 \\
                   5385      & -10770        & 3231      &  7539 
\end{array}
\right)^{-1} 
\end{array}
\right)
\]
\[
-
 21540
\left(
\begin{array}{ccc}
& \\
& \left(
\begin{array}{cccccc}
            8715     &  -12180     \\
            -12180      &  23835       
\end{array}
\right)^{-1} \\
& 
\end{array}
\right).
\]

Having simultaneously both \eqref{eq:local:formula:revisit:example} and \eqref{eq:local:formula:revisit:example:REVERSE} hold is an example of the local algebraic property that we exploit in this article for covariance matrix estimation, and is our main contribution.  

\paragraph{Example: Local inversion for Block Matrices.}
Consider the block matrix
\begin{equation}
\m M =
\left(
\begin{tabular}{ccc}
$\m M_{11}$ & $\m M_{12}$ & $*$\\
 $\m M_{21}$ & $\m M_{22}$ & $\m M_{23}$  \\
 $*$ & $\m M_{32}$ & $\m M_{33}$
\end{tabular}
\right).
\label{eq:block:3x3:matrix}
\end{equation}
Suppose we know that the matrix is invertible and that the inverse has the sparsity pattern
\begin{equation}
 \m M^{-1} =
\left(
\begin{tabular}{rrr}
$\times$ & $\times$ &  $\bm{0}$\\
 $\times$ & $\times$ & $\times$  \\
 $\bm{0}$  & $\times$ & $\times$
\end{tabular}
\right).
\label{eq:block:3x3:matrix:inverse}
\end{equation}
We are not specifying the entry in the top right, nor in the bottom left, of $\m M$, because it can be shown that the sparsity pattern of $ \m M^{-1}$ implies that those two blocks must be
\[
\m M_{13}=\m M_{12}\m M_{22}^{-1}\m M_{23} ,
\]
 and 
 \[
 \m M_{31} = \m M_{32} \m M_{22}^{-1}\m M_{21}.
 \]
There are some mild assumptions, such as that $\m M_{22}$ is invertible.
Then the local inverse formula tells us that:
\begin{equation}
  \m M^{-1} =
\left(
\begin{tabular}{ccc}
$\m M_{11}$ & $\m M_{12}$ & \\
 $\m M_{21}$ & $\m M_{22}$ &  \\
  &  &
\end{tabular}
\right)^{-1}
+
\left(
\begin{tabular}{ccc}
\\
  & $\m M_{22}$ & $\m M_{23}$  \\
  & $\m M_{32}$ & $\m M_{33}$
\end{tabular}
\right)^{-1}
-
\left(
\begin{tabular}{ccc}
\\
& $\m M_{22}$ &  \\
&   &
\end{tabular}
\right) ^{-1}.
\label{eq:block:3x3:matrix:local:inverse:formula}
\end{equation}
Direct matrix multiplication of $\m M$ with this claimed form of $\m M^{-1}$ to arrive at the (block) identity matrix is one way to prove this. Equation \eqref{eq:block:3x3:matrix:local:inverse:formula} is an example of the Local Inverse Formula  \eqref{eq:local:inverse:formula}.
Although this example is only $3 \times 3$, in fact the Local Inverse Formula  \eqref{eq:local:inverse:formula} can be applied to arbitrarily large matrices  (subject to mild assumptions and of course being subordinate to a chordal graph as previously stated), and the proof for larger matrices essentially boils down to this $3 \times 3$ non-commutative block matrix algebra, applied in a recursive way (and the proof is thus by induction), and combined with the key property of chordal graphs that they always have a junction tree \citet{JohnsonLocalInverse1998}.



\subsection{The Local Function}
 Let a chordal graph $\mathcal{C}$ and its junction tree $\mathcal{J}$ be given.
Then we define a function 
$
\f{L}: \mathbb{R}^{n \times n} \longrightarrow \mathbb{R}^{n \times n}
$
by the right hand side of \eqref{eq:local:inverse:formula}
\begin{equation}
\f{L} (\m M) \equiv \sum_{[c] \in \mathcal{C}} \left(\m M_{[c]}\right)^{-1} - \sum_{[j] \in \mathcal{J}} \left(\m M_{[j]}\right)^{-1}.
\label{eq:define:L}
\end{equation}
We now make the following observations:
\begin{enumerate}
\item $\f{L}$ \textit{depends on a chordal graph}, but that dependence is not explicit in the notation on the left side of \eqref{eq:define:L},
i.e., we could plausibly use notation $\f{L}_{\c G}$ to indicate the dependence on the chordal graph $\c G$.
\item The dependence on the chordal graph is seen on the right side of \eqref{eq:define:L} with the first sum being over the maximal cliques $\mathcal{C}$ of the graph, and the second sum being over the separators $\mathcal{J}$. 
This pair $(\mathcal{C},\mathcal{J})$, must correspond to a clique tree (sometimes called a junction tree) of the graph, which has a running intersection property.
It is an equivalence characterisation of chordal graphs that a chordal graph always has at least one clique tree with this property. 
\item At first glance, it also looks like $\f{L}$ depends on the choice of junction tree, but it can be shown that in fact all junction trees would lead to the same final sum on the right of \eqref{eq:define:L}.
\item The domain of $\f{L}$ is a strict subset of $ \mathbb{R}^{n \times n}$.
The input does not need to be an example from our special class of matrices in order to apply $\f{L}$.
The only requirement on the input matrix is that the submatrices on the right hand side of \eqref{eq:define:L} are indeed invertible.
\item The output matrix $\f{L}(\m M)$ is subordinate to the chordal graph, by definition in \eqref{eq:define:L}.
\item For an arbitrary matrix $\m M$, we usually expect $\f{L} (\m M) \ne \m M^{-1}$. However, for matrices for which the inverse is known to be subordinate to a chordal graph, then we have by the results of the local formula that $\f{L} (\m M) = \m M^{-1}$ is indeed the inverse.
\item Recall the doubly sparse constraint in \eqref{eq:naive:constraint}. 
We typically fulfill one of the conditions in that constraint \eqref{eq:naive:constraint} automatically by parameterizing by the allowed nonzero entries of, say, $\m M$.
If the matrix $\m M$ was truly in the special class that satisfied the constraint, then we would have that both 
$\f{L} (\m M) = \m M^{-1}$ and $\f{L} (\m M^{-1}) =\m M$.
We can attempt to exploit this when imposing the constraint, and to take advantage of the property that $\f{L} (\m M)$ or $\f{L} (\m M^{-1})$ will usually be much better to compute than an inverse, because of the local formula.
\end{enumerate}

\subsection{Optimization Summary  and Theoretical Justification}

The following two key points make it clear that the constraints in \eqref{eq:naive:constraint} can be enforced via the local formula, in the way that we describe in our algorithm below.

\begin{theorem}[\cite{BlairPeytonReport1993},   \cite{vandenberghe2015chordal}.]
 Let $\c G$ be a chordal graph, and let  $(\mathcal{C},\mathcal{J})$ be a  corresponding list of cliques and separators, which we call a \emph{perfect elimination ordering}, in terms of a clique tree.
 Then $\m M$ and a Cholesky factor $\m R$ based on this ordering have the same sparsity pattern, in the sense that, if we consider only entries in the triangular part, i.e. $(i<j)$, 
 \[
 \m M_{ij}=0 \qquad \Leftrightarrow  \qquad \m R_{ij} = 0, \qquad (i<j).
 \]
\end{theorem}
One direction of the above result is the `no fill-in' property for perfect eliminators, which has been known a long time in the numerical linear algebra literature (see, e.g., references by Rose and by Rose, Tarjan, \&  Lueker in the exposition of \cite{BlairPeytonReport1993}, or Theorem 9.1 of \citet*{vandenberghe2015chordal}).
The other direction,  is discussed, in relation to \textit{monotone transitivity} properties of chordal graphs, in Figure 4.1, and in equation (9.7)  of \citet*{vandenberghe2015chordal}.

\begin{theorem}[\citet{JohnsonLocalInverse1998}]
  Let $\c G$ be a chordal graph, and let $\f{L}$ be defined as in \eqref{eq:define:L} for this graph $\c G$. 
 Let $\m M$ and $\m M^{-1}$ be a given pair of matrices.
 Assume that $\f{L} (\m M)$, as defined by the right side of equation \eqref{eq:define:L}, is defined.
The following two statements are equivalent.
We have
\[
\m M^{-1} \;\;  \textrm{subordinate to} \;\; \c G
\]
is equivalent to
\[
\m M^{-1} = \f{L} (\m M).
\]
\end{theorem}
One direction of this result is trivial, by definition of our function $\f{L}$.
 The other direction is a result in \citet{JohnsonLocalInverse1998} where it is more general, allowing non-symmetric matrices. 
 More examples, exposition, and references can be found in \citet*{StrangMacNamaraLocalInverseFormula}.



As a consequence of these theoretical results, the optimization \eqref{eq:log:likihood:normal} and \eqref{eq:naive:constraint}, is equivalent to the following. 

Let $\v x \in \mathbb{R}^p$ where $p$ is the number of edges in the given chordal graph $\c G$. 
Let $\m R(\v x)$ be a triangular matrix with a sparsity pattern corresponding to the chordal graph $ \c G$,
i.e., each entry of $\v x$ corresponds to a particular entry of $\m R$.
We will let
\[
\m M(\v x)=\m R(\v x)^\top \m R(\v x).
\]
Let $\m S$ be the sample covariance matrix from $n$ observations.
Then we find the numbers in $\v x$ that \textit{maximize the objective function}
\begin{equation}
\textrm{tr} \left(\f{L} (\m R(\v x)) \f{L} (\m R(\v x))^\top   \m S \right) +  2 \ln\det(\m R(\v x)).
\label{eq:log:likihood:normal:2}
\end{equation}
\textit{subject to the constraint} that 
\begin{equation}
\m C = \m 0
\label{eq:naive:constraint:2}
\end{equation}
where the matrix $\m C$ is defined to be
\[
\m C \equiv \m M \f{L} (\m M) - \m I.
\]
The fact that we can meet the first constraint of \eqref{eq:naive:constraint} in the original optimization problem simply by optimizing over entries with the same sparsity pattern in a Cholesky factor, $\m R(\v x)$, depends on  Theorem 1.
The fact that we can meet the second constraint of \eqref{eq:naive:constraint} in our original problem by requiring $\m C=0$ depends on Theorem 2.
(There is some redundancy in requiring $\m C=\m 0$, since  in our application $\m C$ is symmetric.)

\section{NUMERICAL EXPERIMENTS}
We provide two experiments, one with simulated data, and the other with  financial data obtained from Yahoo Finance. 
\subsection{Simulated Data}
In this experiment, we start with the `true' matrix $\m V_{\mathrm{true}}$ that is given in equation \eqref{eq:first:example:V}.
We reproduce that example matrix here for convenience:
\begin{equation*}
\m V_{\mathrm{true}} =
\left(
\begin{tabular}{cccccc}
    13  &   8  &   4  &   2  &   0 &    0 \\
     8   & 13  &   2  &   1  &  0   &  0 \\
     4   &  2   & 10  &   6  &   1   &  1 \\
     2   &  1   &  6   & 13  &  10  &  10 \\
     0   &  0   &  1   & 10  &  13  &   8 \\
     0   & 0    & 1    & 10  &   8   & 13 
\end{tabular}
\right).
\end{equation*}
We draw $n$ random samples from the multivariate normal distribution with zero mean and with this covariance matrix $\m V_{\mathrm{true}}$, and we form the sample covariance matrix $\m S$.
We then optimize with  Matlab's \texttt{fmincon.m}.
We supply \texttt{fmincon.m} a function handle that it can call to compute the objective function \eqref{eq:log:likihood:normal:2} at a given $\v x$ using the local inverse formula.
The only constraint is that $\m C=\m 0$ as in \eqref{eq:naive:constraint:2}, which is also supplied to \texttt{fmincon.m} as a function handle. We now make the  following observations.

 If $n$ is very large then the optimization returns an estimate that is close to the true covariance matrix.
Also, the likelihood at the estimate is  very close to the likelihood at the true matrix.

However, if $n$ is not large then the estimate can be very noticeably different to the true matrix, and the likelihood at the estimate is higher than the likelihood at the true matrix.
For example, with $n=100$ samples, in one experiment, the sample covariance matrix is
\[
\m S = 
\left(
\begin{tabular}{rrrrrr}
16.703&8.774&4.113&2.629&-0.25&1.16\\
8.774&11.559&1.92&0.01&-1.605&-0.854\\
4.113&1.92&10.07&5.813&1.245&0.947\\
2.629&0.01&5.813&12.424&10.227&9.68\\
-0.25&-1.605&1.245&10.227&13.958&7.88\\
1.16&-0.854&0.947&9.68&7.88&13.345
\end{tabular}
\right),
\]
and the estimate from the optimization procedure is
\[
\left(
\begin{tabular}{rrrrrr}
37.126&16.09&2.384&1.175&0&0\\
16.09&26.676&1.477&0.728&0&0\\
2.384&1.477&10.933&2.507&-2.974&-2.811\\
1.175&0.728&2.507&6.07&4.989&4.715\\
0&0&-2.974&4.989&18.091&5.179\\
0&0&-2.811&4.715&5.179&18.607
\end{tabular}
\right).
\]

\subsection{S\&P100 Financial Data}

\begin{figure}[htb]
\begin{tabular}{cc}
\includegraphics[scale=0.7]{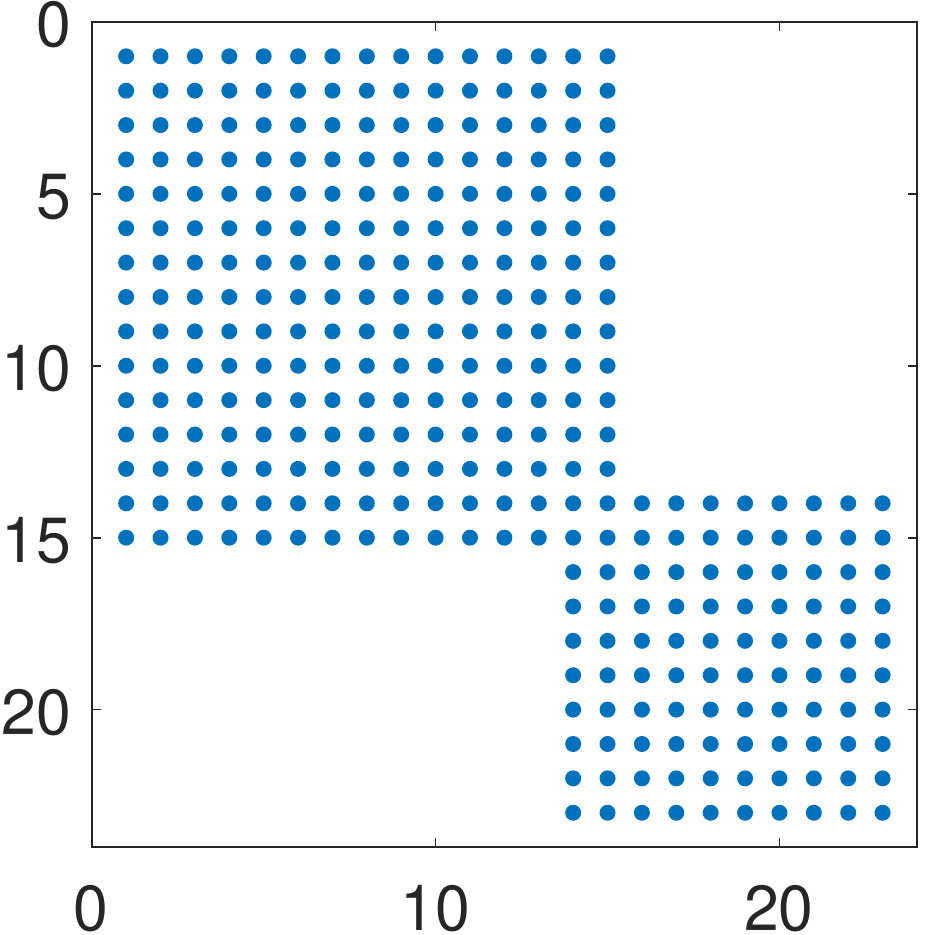} &
\includegraphics[scale=1.1]{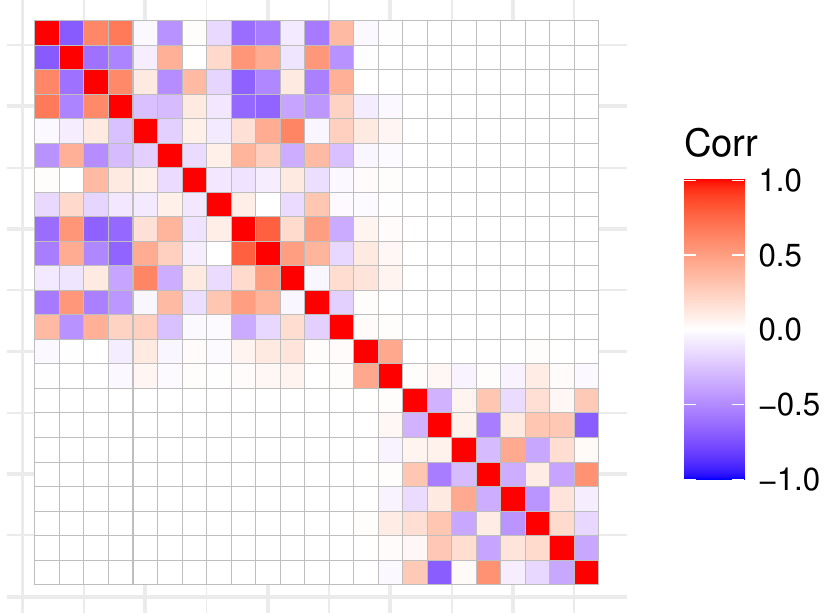}
\end{tabular}
\caption{Left: sparsity pattern of the covariance matrix that is estimated by the method proposed in this article. Note this is NOT simply block diagonal. Right: Visualisation of the corresponding correlation matrix.
Rows 14 and 15 correspond to Amazon and Tesla, while rows 1-13 correspond to `Consumer Discretionary' stocks, and the remaining rows correspond to `Information Technology' stocks, from the  S\&P100 data described in the main text.}
\label{fig:finance:example}
\end{figure}

To further showcase the proposed method of estimation, in Figure~\ref{fig:finance:example}, we considered data based on daily prices of the S\&P 100 stocks from 1 January 2019 to 26 March 2021, downloaded from Yahoo Finance.
In financial markets, portfolio theory suggests nearly all assets have some correlation with common market factors, so perhaps insisting on zeros in the covariance matrix is only justified if we are looking at `residual' covariance matrices, after conditioning on market factors.  
Therefore, we calculated the residuals, after regressing the log returns of all stocks in the S\&P100 on the S\&P100 index.
Then we naively estimated the sample covariance matrix of those residuals. 
For the purpose of demonstration, we chose  the subset of the  S\&P100 index that corresponds to `Consumer Discretionary' (13 stocks) and `Information Technology' (10 stocks), and the corresponding $23 \times 23$ sample covariance matrix.
To apply our proposed method of estimation, we also need a sparsity pattern that corresponds to a chordal graph, so we specified the junction tree to have two cliques, corresponding to the `Consumer Discretionary' stocks, and to the `Information Technology' stocks together with Amazon and Tesla, and specified the separator to correspond to the two stocks Amazon and Tesla.

The result of the procedure is displayed in Figure~\ref{fig:finance:example}.
The examples we used earlier to illustrate this special class of matrices, in \eqref{eq:first:example:V}, were in exact arithmetic, so everything was pleasingly exact and algebraic. 
Here in this application to real data, we obtain our estimates by an optimization procedure, and in finite precision.  
The norm of the constraint, $\| \m M \f{L} (\m M) - \m I \|$ evaluated at the final estimate $\m M$ gives a sense of the accuracy, and in this numerical example the optimization terminated with  $\| \m M \f{L} (\m M) - \m I \| \approx   10^{-4}$.

Note that in the S\&P100 constituents classification, Amazon and Tesla are classified as `Consumer Discretionary,' but arguably these two stocks have more to do with the information technology sector, so visualizing the covariance matrix that we estimate with this sparsity pattern (as in Figure~\ref{fig:finance:example}), could be useful in exploring  the coupling of these two bigger industry groups, via Amazon and Tesla.
In this case the right panel of Figure~\ref{fig:finance:example} shows  the magnitude of the entries in rows corresponding to Amazon and Tesla, and which indicate the strength of  the coupling.

Note also that we could not examine such couplings between groups if we were to only allow simply block diagonal matrices as our estimates, so it is important that the methods are more general, and that is one benefit of allowing the class of chordal graphs. 
All the examples considered here are suggestive of the potential applications of the method we propose.

\subsection{Discussion}
We now highlight a number of issues:
\begin{itemize}
\item
We have chosen a maximum likelihood framework here merely  because it is the simplest way to  illustrate our ideas that estimation  of doubly sparse matrices is indeed possible. 
However, note that the algebraic structures and methods for local computations that we describe do not depend on that maximum likelihood framework -- so, for example, it should be possible to also use these computational approaches in other frameworks that do not make assumptions about the distribution.

\item
The choice of norm is likely important (for example, the $1-$norm may be preferable to the $2-$norm), but that issue is not explored here.

\item
Instead of prescribing the sparsity pattern, it would be better to estimate the sparsity graph from the data.
(For instance, in the finance example, we would prefer not to require any hunch about the significance of Amazon or Tesla.)
A crude first approach could be in three separate steps: first use existing methods such as the graphical LASSO to estimate a graph, and then second force the resulting graph to be chordal, and then third and finally apply the suggested method of this paper. 
But instead of such a crude approach of consecutive but separate estimation problems, it seems more natural to simultaneously estimate  the graph together with the matrix estimation problem.

\item We have required the matrix and its inverse be subordinate to the same chordal graph, because that it is the simplest first idea to explore.
Note that -- although we have not displayed such an example pair of matrices in this article -- if the $(i,j)$ edge is present in the graph, then it is possible that the $(i,j)$ entry of $\m M$ is nonzero and the $(i,j)$ entry of $\m M^{-1}$ is zero, and that both matrices are subordinate to the same chordal graph.
 It  is natural to also consider the generalisation of our problem to the case where  the matrix and the inverse are subordinate to different chordal graphs, but we did not explore that generalisation here.

\item If the matrix $\m M$ was truly in the special doubly sparse class, then  the matrix would be a fixed point of the special Local Function we defined in \eqref{eq:define:L}, composed with itself, i.e. $\f{L} ( \f{L} (\m M) )= \m M$. 
We have not explored fixed-point iteration algorithms.
Nor have we explored the algebraic structure of this special class of matrices from this point of view of the properties of the local function $\f{L} (\cdot)$, although The Nullity Theorem gives constraints on ranks of subblocks \cite{strang2004interplay}.

\item Chordal graphs can be considerably more complicated than only the simplest examples we have illustrated here -- see many varied examples of matrix sparsity patterns and corresponding graphs in  the references, e.g., \citet*{vandenberghe2015chordal}.

\item Any given graph  can be `forced' to become a chordal graph by a known process of adding edges, thus allowing the methods of this paper to always be used.  
However, whether or not such a process is computationally worthwhile   depends on the example. 

\end{itemize}

\section{CONCLUSION}
We proposed a method to estimate a covariance matrix when both the covariance and the precision matrix are sparse (which we called double sparsity).
This is a maximum likelihood approach, subject to the double sparsity  constraint.   
This appears to be  the first work to estimate such special pairs of covariance and precision matrices.
The sparsity patterns we consider are restricted to the class of chordal graphs (also known as decomposable graphs in the graphical models literature). 
This class includes the banded matrices with banded inverse.
Restricting to this class of sparsity pattern allows us to exploit a special algebraic structure --  the local inverse formula, as we described -- that can make computations faster (and that is computationally more attractive than simply naively imposing the double sparsity constraint during any optimization).
For future work, it should be possible to extend these approaches to simultaneously estimate both the sparsity pattern, and the corresponding special pair of covariance matrix and precision matrix.


\bibliographystyle{plainnat}
\bibliography{shevCCIrefs}

\begin{thebibliography}{21}
\providecommand{\natexlab}[1]{#1}
\providecommand{\url}[1]{\texttt{#1}}
\expandafter\ifx\csname urlstyle\endcsname\relax
  \providecommand{\doi}[1]{doi: #1}\else
  \providecommand{\doi}{doi: \begingroup \urlstyle{rm}\Url}\fi

\bibitem[Bickel and Levina(2008)]{bickel2008regularized}
Peter~J Bickel and Elizaveta Levina.
\newblock Regularized estimation of large covariance matrices.
\newblock \emph{The Annals of Statistics}, 36:\penalty0 199--227, 2008.

\bibitem[Blair and Peyton(1993)]{BlairPeytonReport1993}
J.R.S. Blair and B~Peyton.
\newblock Introduction to chordal graphs and clique trees.
\newblock In \emph{Graph Theory and Sparse Matrix Computation}. Springer, 1993.

\bibitem[Dempster(1972)]{Dempster72}
A.P. Dempster.
\newblock Covariance selection.
\newblock \emph{Biometrics}, 28\penalty0 (1):\penalty0 157--175, 1972.

\bibitem[Dym and Gohberg(1981)]{dym1981extensions}
Harry Dym and Israel Gohberg.
\newblock Extensions of band matrices with band inverses.
\newblock \emph{Linear algebra and its applications}, 36:\penalty0 1--24, 1981.

\bibitem[Friedman et~al.(2008)Friedman, Hastie, and
  Tibshirani]{friedman2008sparse}
J.~Friedman, T.~Hastie, and R.~Tibshirani.
\newblock Sparse inverse covariance estimation with the graphical lasso.
\newblock \emph{Biostatistics}, 9\penalty0 (3):\penalty0 432--41, 2008.

\bibitem[Georgescu et~al.(2018)Georgescu, Higham, and
  Peters]{georgescu2018explicit}
Dan~I Georgescu, Nicholas~J Higham, and Gareth~W Peters.
\newblock Explicit solutions to correlation matrix completion problems, with an
  application to risk management and insurance.
\newblock \emph{Royal Society open science}, 5\penalty0 (3):\penalty0 172348,
  2018.

\bibitem[Golub and Pereyra(2003)]{golub2003separable}
Gene Golub and Victor Pereyra.
\newblock Separable nonlinear least squares: the variable projection method and
  its applications.
\newblock \emph{Inverse problems}, 19\penalty0 (2):\penalty0 R1, 2003.

\bibitem[Grone et~al.(1984)Grone, Johnson, S{\'a}, and
  Wolkowicz]{grone1984positive}
Robert Grone, Charles~R Johnson, Eduardo~M S{\'a}, and Henry Wolkowicz.
\newblock Positive definite completions of partial hermitian matrices.
\newblock \emph{Linear algebra and its applications}, 58:\penalty0 109--124,
  1984.

\bibitem[Johnson and Lundquist(1998)]{JohnsonLocalInverse1998}
C.R. Johnson and M.~Lundquist.
\newblock Local inversion of matrices with sparse inverses.
\newblock \emph{Linear Algebra and Its Applications}, 277:\penalty0 33--39,
  1998.

\bibitem[Lauritzen(1996)]{GraphicalModelsLauritzenBook}
S.~Lauritzen.
\newblock \emph{Graphical Models}.
\newblock Oxford University Press, 1996.

\bibitem[Ledoit and Wolf(2004)]{ledoit2004honey}
Olivier Ledoit and Michael Wolf.
\newblock Honey, i shrunk the sample covariance matrix.
\newblock \emph{The Journal of Portfolio Management}, 30\penalty0 (4):\penalty0
  110--119, 2004.

\bibitem[Pourahmadi(2013)]{PourahmadiCovarianceMonograph}
Mohsen Pourahmadi.
\newblock \emph{High-dimensional covariance estimation}.
\newblock Wiley, 2013.

\bibitem[Ravikumar et~al.(2011)Ravikumar, Wainwright, Raskutti, and
  Yu]{ravikumar2011high}
Pradeep Ravikumar, Martin~J Wainwright, Garvesh Raskutti, and Bin Yu.
\newblock High-dimensional covariance estimation by minimizing l1-penalized
  log-determinant divergence.
\newblock \emph{Electronic Journal of Statistics}, 5:\penalty0 935--980, 2011.

\bibitem[Rothman et~al.(2010)Rothman, Levina, and Zhu]{Rothman2010}
A.~J. Rothman, E.~Levina, and J.~Zhu.
\newblock A new approach to cholesky-based covariance regularization in high
  dimensions.
\newblock \emph{Biometrika}, 97:\penalty0 539--550, 2010.

\bibitem[Salemi et~al.(2019)Salemi, Song, Nelson, and Staum]{l2019gaussian}
Peter Salemi, Eunhye Song, Barry Nelson, and Jeremy Staum.
\newblock Gaussian markov random fields for discrete optimization via
  simulation: Framework and algorithms.
\newblock \emph{Operations Research}, 67\penalty0 (1):\penalty0 250--266, 2019.

\bibitem[Speed and Kiiveri(1986)]{SpeedKiiveri1986}
T.P. Speed and H.T. Kiiveri.
\newblock Gaussian {M}arkov distributions over finite graphs.
\newblock \emph{The Annals of Statistics}, 14:\penalty0 138--150, 1986.

\bibitem[Strang(2010)]{strang2010fast}
Gilbert Strang.
\newblock Fast transforms: Banded matrices with banded inverses.
\newblock \emph{Proceedings of the National Academy of Sciences}, 107\penalty0
  (28):\penalty0 12413--12416, 2010.

\bibitem[Strang and MacNamara(2018)]{StrangMacNamaraLocalInverseFormula}
Gilbert Strang and Shev MacNamara.
\newblock A local inverse formula and a factorization.
\newblock In Woźniakowski~H. Dick~J., Kuo~F., editor, \emph{Contemporary
  Computational Mathematics - A Celebration of the 80th Birthday of Ian Sloan},
  pages 1109--1126. Springer, 2018.
\newblock \doi{10.1007/978-3-319-72456-0_51}.
\newblock URL \url{arXiv:1610.01230}.

\bibitem[Strang and Nguyen(2004)]{strang2004interplay}
Gilbert Strang and Tri Nguyen.
\newblock The interplay of ranks of submatrices.
\newblock \emph{SIAM review}, 46\penalty0 (4):\penalty0 637--646, 2004.

\bibitem[Vandebril et~al.(2007)Vandebril, van Barel, and
  Mastronardi]{SemiSeparableBook}
R.~Vandebril, M.~van Barel, and N.~Mastronardi.
\newblock \emph{Matrix Computations and Semiseparable Matrices}.
\newblock Johns Hopkins, 2007.

\bibitem[Vandenberghe and Andersen(2015)]{vandenberghe2015chordal}
Lieven Vandenberghe and Martin~S Andersen.
\newblock Chordal graphs and semidefinite optimization.
\newblock \emph{Foundations and Trends in Optimization}, 1\penalty0
  (4):\penalty0 241--433, 2015.

\end{thebibliography}

\section*{AUTHOR BIOGRAPHIES}

\noindent {\bf Shev MacNamara} is a Senior Lecturer in the School of Mathematical and Physical Sciences at the
University of Technology Sydney. His research interests include advanced stochastic modeling
and simulation. Previously, he was a postdoctoral scholar in the Department of Mathematics at MIT, and
a postdoctoral scholar in the Mathematical Institute at The University of Oxford. He has been a Fulbright
Scholar, and has held a David G. Crighton Fellowship at The University of Cambridge. His webpage is
\url{https://www.uts.edu.au/staff/shev.macnamara}, and his email is \texttt{shev.macnamara@uts.edu.au}.
  \\

 \noindent {\bf  Erik Schl\"ogl} is Professor and Director of the Quantitative Finance Research Centre, University of Technology Sydney, Broadway, NSW 2007, Australia.  He also holds an honorary Professorship at the African Institute for Financial Markets and Risk Management (AIFMRM), University of Cape Town, Rondebosch 7701, South Africa; and an honorary affiliation with the Department of Statistics, Faculty of Science, University of Johannesburg, Auckland Park 2006, South Africa.  His email address is \texttt{Erik.Schlogl@uts.edu.au}. His website is \url{https://profiles.uts.edu.au/Erik.Schlogl}.\\
	
\noindent {\bf Zdravko Botev}  is a Lecturer of Statistics at UNSW Sydney. His research interest include: 1) Monte Carlo variance reduction methods, especially for rare-event probability estimation; 2) nonparametric kernel density estimation, and more recently 3) fast model-selection algorithms for  large-scale statistical learning. He is well-known as the inventor of the widely-used method of \emph{kernel density estimation via diffusion}, as well as  the \emph{generalized splitting algorithm} for Monte Carlo rare-event simulation and combinatorial counting.   
His website is \url{https://web.maths.unsw.edu.au/~zdravkobotev/} and email address is \texttt{botev@unsw.edu.au}.

\end{document}